\documentclass[11pt,preprint2]{aastex}

\begin{document}

\title{X-ray Outburst in Mira A}
\author{Margarita Karovska\altaffilmark{1}, Eric
Schlegel\altaffilmark{1}, Warren Hack\altaffilmark{2},
John Raymond\altaffilmark{1}, and Brian E. Wood\altaffilmark{3}}

\affil{Harvard-Smithsonian Center for Astrophysics,60 Garden Street,
Cambridge, MA}
\altaffiltext{1}{Harvard-Smithsonian Center for Astrophysics,60 Garden Street, Cambridge, MA}

\altaffiltext{2}{Space Telescope Science Institute, Baltimore, MD 21218.}

\altaffiltext{3}{JILA, University of Colorado and NIST, Boulder, CO
  80309-0440.}

\begin{abstract}

We report here the Chandra ACIS-S detection of a bright
soft X-ray transient in the Mira AB interacting symbiotic-like binary.
We resolved the system for the first time in
the X-rays. Using Chandra and HST images we determined that the
unprecedented outburst is likely associated with the cool AGB star
(Mira A), the prototype of Mira-type variables. X-rays have never
before been detected from an AGB star, and the recent activity signals
that the system is undergoing dramatic changes. The total
X-ray luminosity of the system is several 
times higher
than the luminosity estimated using previous XMM and ROSAT
observations.  The outburst may be caused by a giant flare in Mira A
associated with a mass ejection or a jet, and may have long term consequences
on the system.

\end{abstract}


\section{Introduction}

Mira AB is an interacting binary system composed of an aging cool
giant (Mira A) losing mass at a rate of $\sim$$10^{-7}$ ${M_{sol}}/yr$
(Bowers \& Knapp, 1988), and an accreting companion (Mira B) about
$\sim$70~AU ($\sim$0.6'') away.  This detached binary is one of very
few wind accreting systems that has been spatially resolved and for
which the energy distribution of both components can be determined
{\it unambiguously} (Karovska {\it et al.}, 1997).  Studies of Mira AB
wind accretion and mass transfer at wavelengths ranging from X-rays to
radio provide a basis for understanding wind accretion processes in
many other astronomical systems that currently cannot be resolved.
Therefore this nearest symbiotic-like system offers a test-bed for
detailed studies of wind accretion processes and accretion theory
(Livio 1988).

In 1995, using the HST FOC, we resolved the system for the first time spatially and
spectrally at UV and optical wavelengths and studied the interacting
components individually (Karovska {\it et al.} 1997). 
In the past few
years we have been witnessing changes in the spectral
energy distribution (SED) of Mira AB, especially 
in UV
 (Karovska
{\it et al.} 1997; Wood, Karovska, \&Hack 2001). 
In addition to the general fading of the accretion luminosity, another
baffling 
development was the appearance of a forest of
Ly$\alpha$-fluoresced H$_{2}$ emission lines, which dominated the HST
spectra in 1999, despite not being seen at all in the 1995
spectra or by IUE (Wood, Karovska, \& Hack 2001; Wood, Karovska,
\& Raymond, 2002).
A similar drop in the accretion luminosity and appearance of a set of
Ly$\alpha$-fluoresced H$_{2}$ emission lines were also
seen in the FUSE spectra in 2001 (Wood \& Karovska 2004). 
Variable mass loss from Mira A
could have caused the 
changes in the UV flux of Mira B
observed in the past 10 years
signaling that the system is
undergoing dramatic changes.

In 1993, the {\it ROSAT}
observation of Mira AB resulted in the first unambiguous detection of
X-ray emission from Mira AB (Karovska at al. 1996). 
This observation
resolved the contradicting results from
the analysis of the {\it EINSTEIN} observation; Jura \&
Helfand (1984) marginally detected an X-ray source, while Maggio
et al. (1990) set an upper limit of
$f{_x}$$<$1.4 $\times$10$^{-13}$ erg s$^{-1}$ cm$^{-2}$.
The {\it ROSAT} X-ray luminosity of the Mira AB system was estimated 
$\sim$10$^{29}$erg s$^{-1}$ (Karovska et al. 1996) which is similar to
the luminosity estimated from the XMM observations carried out about
ten years later (Kastner \& Soker 2004b).

In this {\it Letter} we report 
the initial results from our recent
Chandra observations 
and from coordinated HST and ground-based observations of Mira AB,
including our discovery of a soft X-ray outburst in
Mira A.

\section{Observations and Analysis}

On 2003 December 6 we carried out a 70 Ksec pointed Chandra
observation
of Mira AB using the ACIS-S instrument (Weisskopf {\it et al.} 2002).
The source was placed at the nominal aim point of ACIS-S, on CCD S3.
We analyzed the {\it Chandra} observations using {\it CIAO} data
reduction and analysis routines.
\footnote{CIAO is the Chandra Interactive Analysis of Observation's
data analyses system package (http://cxc.harvard.edu/ciao)}
The source spectrum was extracted using an aperture of 12$''$ radius
centered on Mira AB; the background came from a surrounding annulus of
inner radius 14$''$ and outer radius 30$''$.  
Figure 1 shows the Chandra spectrum and the best spectral fit
(described in detail in \S3.2).
The source contained
5115 counts for a count rate of 0.073$\pm$0.001 cts s$^{-1}$.  The
background rate was 0.0007 cts s$^{-1}$.  In the 0.2-0.7 keV band, the
count rate was 0.066$\pm$0.001 cts s$^{-1}$ (background: 0.0003 cts
s$^{-1}$); in the 0.7-4 keV band, the detected rate was 0.008 cts
s$^{-1}$ (background: 0.0004 cts s$^{-1}$).  The combination of the
declining effective area below 0.5 keV plus the excess absorption
layer (Chandra Proposer's Guide\footnote{http://asc.harvard.edu/proposer/POG/index.html}) ensure minimal or zero photon
pile-up.  Pushing all of the model parameters to their most
pile-up-favorable extremes yields a pileup estimate of less than 4\%.

We detected several thousand counts below 1 keV associated with a new bright soft
source in the system, which was not reported a
few months before by XMM (Kastner \& Soker, 2004b), or by ROSAT in
1993 (Karovska et al., 1996). 
The high-energy component ($>$1keV) is similar in appearance to the quiescent
XMM and ROSAT spectra. However, a detailed comparison shows
that a clear evolution has occurred in this portion of the spectrum
($\sim$1.2-1.8 keV) as well.
The count rates (see below), if folded through the XMM or ROSAT effective areas, would have been easily detected in either instrument.  Furthermore,
the spectrum is dominated by the soft component in the 0.2-0.7 keV
band and the spectral behavior of the ROSAT and XMM spectra is
significantly harder.

Following the Chandra X-ray outburst detection, we requested and
obtained Director's time on Chandra and HST.  Chandra HRC/LETG
observations (40 Ksec) were carried about two months after the initial
Chandra observations (on January 11 2004). We detected signal only in the
zeroth order of the spectrum which showed that the brightness of the
soft source had dropped significantly.  This observation detected
$\sim$80 counts.  Were the soft ACIS spectral component (\S3.2) still present
with the same absorbed flux at the time of the LETG observation, $>$250 counts would have been
detected in the zero-order image. However, the detected counts could
be due to the hard source as well, or to both sources. Further
modeling using the HRC and the ACIS observations is in progress.

We carried out follow-up HST STIS CCD and MAMA observations about two
months after the initial Chandra observations (2004 February 2 and
February 16).  We resolved the system in the near-UV and obtained
individual spectra of both components.  MultiDrizzle techniques
(Koekemoer at al. 2002) were used to combine multiple CCD
images recorded using the F28x50OII filter centered on the [OII] 3729
{\AA} line. The images resolved the system and detected extended
emission.  The STIS spectra show a significant increase in the Mg
h \& k lines emission in Mira A and in Mira B when compared to the
1999 STIS observations. Velocity shifts of the order of 100 km/s were
also detected in Mira A's Mg h \& k lines.  A similar increase of line
emission was detected in the optical spectra ({H$_\alpha$} and
OIII$\lambda$5007$\AA$ lines) obtained at the Oak Ridge
observatory. In addition, broadening and complex structure were
observed in the {H$_\alpha$} during the month following the
X-ray outburst, indicating velocities of $\sim$ 100 km/s.  
Further details of the analysis and the results of the Chandra, HST, and
ground-based observations will be discussed elsewhere (Karovska {\it et al.}, 2005, in prep.).

\section{Results}

\subsection{X-ray and near-UV Imaging}

The ACIS-S raw image of Mira AB binned at
 the detector 0.495'' pixel
resolution showed an elongation along the Mira AB system axis (Figure
 2a).
We made soft ($<$ 0.7 keV) and hard (0.7-2 keV) images of Mira AB, 
filtered based on the major components detected in the ACIS-S spectrum.
Figure 2b shows
the image of the hard source (to the East) with an overlay of the contours of the soft
 component (to the West).
The centroids of these sources
are displaced by $\sim$ 0.6''. 

We attempted to do further spatial analysis using the HRC zeroth order
image.
The image appears extended and it is shifted $\sim$0.3 arcseconds 
from the soft ACIS image, in the direction toward the hard ACIS image. 
It could therefore be associated with Mira A or Mira B (or both). 
However, the low counts and lack of spectral resolution make it difficult to
determine if there is a component associated with a remnant of the outburst.
Although the HRC pixels ($\sim$0.14") are indeed smaller than the ACIS
pixels, the actual resolution of the HRC is in fact comparable to ACIS.
Furthermore, there are additional artifacts in the HRC images that makes it 
very difficult deciding if the extension in the image to the South and
an additional source ~0.5" to the West are real.

We explored the spatial
extent of the emission in the ACIS-S (0.3-4 keV) image of the system at 0.2'' 
resolution using
a new multiscale
deconvolution technique {\it EMC2} (Esch {\it et al.} 2004)
and {\it Chandra} PSF model.
This was possible because the 
{\it Chandra} data include information about the photon energies and
positions which was used to obtain filtered images and carry out
sub-pixel resolution analysis.  
The Chandra PSF varies as a function of energy and
 off-axis angle; we carried out PSF simulations using 
CIAO software and threads, including the ChaRT PSF
simulator\footnote{http://asc.harvard.edu/chart/index.html)}. The PSF
simulation was carried out using information on the spectral
distribution and off-axis location of the system as inputs to ChaRT.
The {\it EMC2} deconvolution
technique, described in detail in Esch {\it et al.} 2004, 
was developed for low-count statistics data,
and it provides error estimates in addition to the reconstructed
images.

The deconvolved image (color image in Figure 3) shows two sources
separated by $\sim$0.6''. The pixel size is 0.1'' (0.2 ACIS-S pixel size).
This is the first image of an interacting binary that has been
spatially resolved at X-ray wavelengths.
We note that the location of the
brighter source in the deconvolved image (to the West) corresponds to 
the
position of the soft source (as determined using filtered ACIS-S
images), and the fainter source (to the East) corresponds to
the hard-band ACIS-S image.

We compared the X-ray images with the
HST images of the Mira AB components obtained two months
later, showing two sources separated by $\sim$0.6''. In the HST images 
Mira A is to the west of Mira B.
Figure 3 displays the overlay of the contours of the HST 3729 $\AA$  
image on the Chandra
deconvolved image, showing that the soft X-ray source 
is in the vicinity of the
of Mira A. The soft source is therefore likely associated with the AGB
star rather
than with the accreting companion Mira B.
When the X-ray image is compared to the HST image of Mira A and Mira B, we see 
slightly larger separation of the X-ray components ($\sim$0.1'') and possible
small rotation of the axis between the sources. This difference could be real,
but also it could be due to uncertainties in the
deconvolution, or to the limited resolution in the Chandra data.
It could also be a result of HST absolute positioning uncertainties. 

The X-ray
images of Mira A and Mira B appear extended.
Furthermore, the Chandra image shows a faint ``bridge-like'' feature
extending  between the components. 
Similar ``bridge-like'' structure was detected
in the 1995 HST images indicating possible mass flow between 
the components (Karovska et al. 1997). 
Although the features in the X-ray image are uncertain 
because of the low
count level and/or possible PSF and deconvolution artifacts, we note
that very similar structures can be seen in the overlayed HST
contours.

We determined that some of the extended structure in the HST image to the
right of Mira A, in the
opposite direction from Mira B, is
due to an undocumented red leak of the F28x50OII filter. Following the HST
observations we discovered and confirmed the red leak
in other very red sources.  For example,  F28x50OII filter images of
CH Cyg, which contain a very red component, show similar
structure, and we have detected similar structured PSF in the
[OIII]5007 filter during the 1999 observation of Mira AB.

The evidence of extended structures in both components in both X-ray
and UV images is
uncertain and require further analysis and modeling.
In order to determine if there was a mass ejection
during the outburst, it will be
necessary to carry out further
high-angular resolution monitoring including at X-ray and UV wavelengths.

\subsection{X-ray Spectroscopy}

The spectrum was binned to a minimum of 20 counts per channel.  The
effective area was corrected for the energy- and time-dependent
instrumental absorption and {\tt Xspec} v11.0 was used to fit the spectra (Arnaud
1996) as detailed below.
The X-ray spectrum shows strong soft emission and potentially
multiple harder components.  Standard continuum models
(bremsstrahlung, blackbody, power law) did not provide good fits to
the soft component.  We describe the spectral fit below with
90\% errors on the model parameters.

We fit the low-energy portion of the spectrum using an absorbed model
spectrum consisting of a sum of
gaussians, each with a fixed center and zero width to represent an
un-resolved line. We included several C, N, and O lines. The 
line centers (in keV) are at C V 0.299,
0.304, 0.308; C VI 0.367; N VI 0.319, 0.426, 0.431; C VI 0.435, 0.459;
N VII 0.500; O VII 0.561, 0.569, 0.574; and O VIII 0.653).  For
each line, only the normalization was fit.  The initial fit showed
that the  O line normalizations were all consistent with zero as were
approximately half of the C and N normalizations; a subsequent fit
used a model containing only the C and N lines.  Figure 1 shows the
result with lines at C VI 0.367 keV, N VI 0.426 keV, C VI 0.459 keV, and N VII
0.500 keV (line normalizations = 4.9, 6.2, 50.8, and 33.1,
respectively, in units of 10$^{-5}$ photons cm$^{-2}$ s$^{-1}$).  
Because of the limited spectral resolution we cannot determine if
there are any specific lines detected, or if the fit is unique.
When we
added a continuum component, its model normalization was consistent
with zero.  The apparently large residuals near 0.3 keV may be
explained by the known problems with the corrections for the
contaminant near the carbon edge (Marshall 2003\footnote{Available at
http://space.mit.edu/ASC/calib/letg\_acis/ck\_cal.html.}).  The best-fit
spectrum is therefore most easily explained as blended emission of C +
N lines.

The hard component was best-fit with a bremsstrahlung continuum of kT
$\sim$0.78$^{+0.15}_{-0.23}$ keV to mimic an optically thin thermal
plasma, plus two zero-width gaussians representing emission lines at
1.02$^{+0.06}_{-0.04}$ and 1.36$^{+0.03}_{-0.04}$ keV, all absorbed by
a column N$_{\rm H} \sim$9.0$^{+0.15}_{-0.12}{\times}$10$^{22}$
cm$^{-2}$.  Other models provided very poor fits including models
containing additional gaussians at the positions of expected line
emission.  
The fitted bremsstrahlung temperature is nearly identical with the
value determined from the XMM spectrum (Kastner \& Soker 2004b).  The
fitted column is about a factor of two higher than the XMM-determined
value.  A specific error range is not shown for the XMM value, but the
authors state ``formal uncertainties are $\approx$20\%''.  A 20\%
error range on the XMM value places it outside the 90\% range of the
fitted value determined here suggesting the column may have increased
by a factor of $\approx$2 between the two observations.
The hard gaussians have equivalent widths of 236$^{+244}_{-110}$ and
124$^{+124}_{-50}$ eV, respectively, and we attribute them to Ne and
Mg emission lines. We note, Ne IX and Ne X lines were also
detected by the XMM (Kastner \& Soker 2004), but the Mg
line is not pronounced in the XMM spectrum. 
These results are similar to the enhanced K$\alpha$ H-like Ne and
He-like Mg lines
observed in the ASCA spectrum of the symbiotic CH Cyg showing jet
activity (Ezuka et
al. 2001), corresponding to log(T)=6.8. Our signal-to-noise is
insufficient to fit for the He-like Si line (found in the 
CH Cyg spectrum). These high temperature lines may indicate shock
heated emission or could be associated with the flare or a jet-like ejecta.

We calculated the total flux from the system 5.6$\times$10$^{-13}$ (0.2-4
keV), 5.5$\times$10$^{-13}$ (soft component only in 0.2-0.7 keV), and
2.7$\times$10$^{-14}$ (hard component only in 0.7-4 keV) erg s$^{-1}$
cm$^{-2}$ for the bands defined.
The unabsorbed fluxes are very sensitive to the adopted column.  The
fitted column toward the soft source is $\sim$8.0$\times$10$^{19}$
cm$^{-2}$.  This value, if adopted as the column {\it toward the
system}, is not consistent with the column obtained from the ROSAT
spectrum (Karovska et al. 1996) nor the XMM observation (Kastner \&
Soker 2004b).  However, the fitted column is really only a lower limit.
The 90\% confidence contour is closer to
$\sim$2.8${\times}$10$^{20}$ cm$^{-2}$.  Within this upper
limit, the column remains inconsistent with the XMM value, but is consistent
with the column determined from the UV H$_{\rm 2}$ line spectroscopy
(Wood et al. 2002).
If we adopt the formal-fit column from the soft spectrum, 
then the unabsorbed fluxes
are 6.4$\times$10$^{-13}$ (0.2-4 keV), 6.2$\times$10$^{-13}$ (0.2-0.7
keV), and 2.7$\times$10$^{-14}$ (0.7-4 keV) erg s$^{-1}$ cm$^{-2}$.
If we adopt the upper limit value for the column, the unabsorbed
fluxes are 8.3$\times$10$^{-13}$ (0.2-4 keV), 8.0$\times$10$^{-13}$
(0.2-0.7 keV), and 2.7$\times$10$^{-14}$ (0.7-4 keV) erg s$^{-1}$
cm$^{-2}$.
Adopting the fitted value of the column for the hard component, the
unabsorbed hard band flux is 9.9$\times$10$^{-14}$ erg s$^{-2}$
cm$^{-2}$ (0.7-4 keV).

We emphasize that the appropriate N$_{\rm H}$ value could be
completely different for the hard and soft sources since the
emissions originate from different places.  It is conceivable that a
persistent low energy X-ray source exists, and that it was made
visible by a temporary drop in the neutral column density surrounding
Mira A.  It seems more likely, however, that an X-ray flare ionized
its surroundings, producing the formal value of smaller column as a
side effect.
Taking note of the different columns toward the sources, the
unabsorbed luminosity then falls into the range of
1.5-2$\times$10$^{30}$ erg s$^{-1}$, for an assumed distance of 130
pc, and assuming the luminosity is the sum of the soft and hard
components separately.  The range results from the adopted value of
the column toward the soft component.  The total luminosity is about
a factor of 2-4 higher than that observed by XMM and ROSAT.
A detailed analysis and modeling of the {\it Chandra} data
combined with the multi-wavelength
observations of the outburst will be discussed
elsewhere (Karovska {\it et al.}, 2005, in prep.).

\section{Possible Causes and Consequences of the X-ray Outburst}

Detecting X-ray emission from Mira A is 
a very important result because X-ray emission has not
been detected so far in single AGB stars, including Miras (e.g.
Kastner \& Soker 2004a). 
The soft X-ray outburst in Mira A could be caused
by a magnetic flare followed by a large mass ejection, analogous to the solar flares and Coronal Mass
Ejections(CMEs). 
Soker and Kastner (2003) suggested that 
magnetic reconnection events near the stellar surface should lead to 
localized, long-duration flares.
The crude time scale for a flare might be estimated
by multiplying the 1-2 day time scale for a large flare on
an RS CVn star by the ratio of Mira A's radius (few hundred solar radii) to   
that of the RS CVn star (4-15 solar radii), suggesting time scales
from weeks to months, which is in accordance with the time scale of
the X-ray event observed by Chandra.

In the case of mass ejection we could  expect
changes in the SED of both components on a time scale of months to
years. As the ejected material cools down we expect a significant
increase in dust
formation in the system in 2004/2005.
This is also the time scale on which we would expect
a response of the accretion disk around Mira B to any disturbance
caused by an outburst on Mira A in late 2003, if the flow is
propagating toward Mira B with speed of $\sim$ hundred km/s (as
suggested by our preliminary analysis of the HST and ground-based
spectroscopy). 
An increased outflow from Mira A resulting from the recent outburst 
could help renew the accretion
disk 
around
the companion, following the low-state observed by HST and
FUSE in 1999--2001.

The increased accretion rate into Mira B could also cause instabilities in the
accretion 
disk and jet-like activity (e.g. Soker 2002) similar to that detected in several nearby unresolved
symbiotic systems (e.g. in CH Cyg, Corradi et al. 2001, R Aqr,
Kellogg et al. 2001).
An outburst in Mira B associated with 
the accretion disk
instabilities could be analogous to a dwarf nova outburst except
that the much larger size of the wind-fed disk in
this well-separated binary would increase the
duration from about a week to many months.  
Further multiwavelength observations are necessary
to determine the nature and characteristics of
 instabilities in the system and/or its individual components that
may have caused the outburst, and to understand the long-term consequences of
this outburst on
the system.

\acknowledgements

We are grateful to Drs H. Tananbaum and S. Beckwith
for granting Director's time on Chandra and HST.
This work is part of a long term
collaboration with several colleagues on coordinated multi-wavelength
campaign including Drs. R. Stefanik,
M. Marengo, J. Drake, A. Henden, D. Esch, L. Matthews, E. Guinan, 
E. Waagen
and the AAVSO. We dedicate this paper to
Janet A. Mattei who has inspired
this work and made these observations possible for many years.
We thank the referees N. Soker and
J. Kastner for helpful suggestions.
This work was supported by NASA grants GO4-5024A and NAS8-39073.
MK and ES are members of the Chandra X-ray Center, which is operated
by the Smithsonian Astrophysical Observatory under contract to NASA NAS8-39073.

\newpage

\begin{figure}[t]
\centering
\scalebox{0.6}{\rotatebox{-90}{\includegraphics{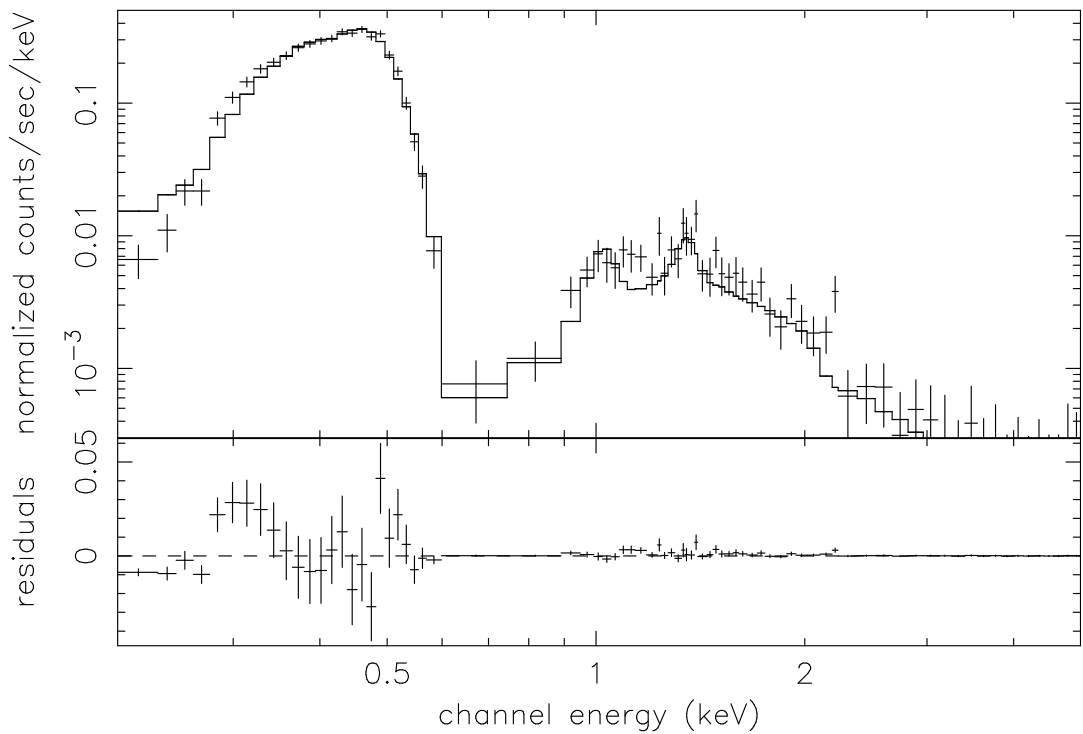}}}
\caption{{\it Chandra} ACIS-S spectrum of Mira AB (plus signs) fit
  with a combination of gaussians for the soft spectral component and
  a bremsstrahlung plus gaussians for the hard component (see  \S3.2).  Residuals
  fall mostly near the C edge at 0.3 keV where the response matrix is
  known to have errors.}
\label{mira_spec}
\rule{3.0in}{0.2mm}
\end{figure}

\clearpage

\newpage

\begin{figure}[t]
\centering
\scalebox{0.830}{\rotatebox{0}{\includegraphics{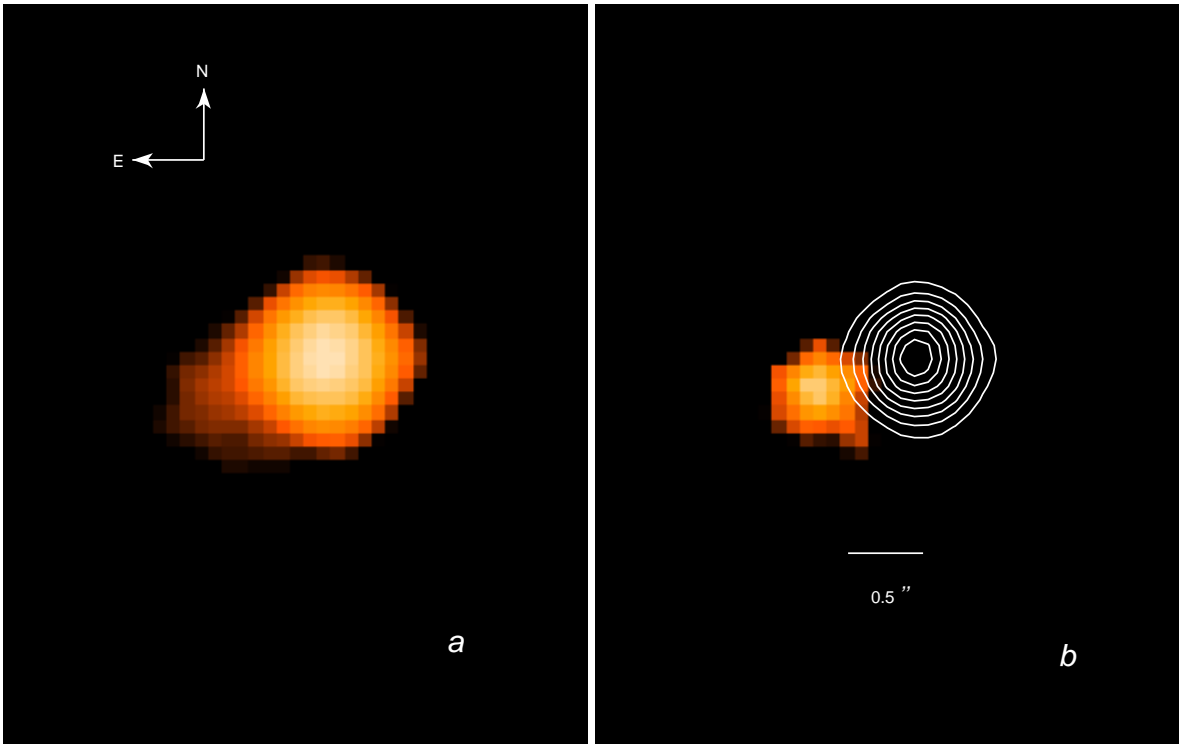}}}
\caption{{\it Chandra} images of Mira AB: (a) ACIS-S raw image of
Mira AB filtered from 0.3 to 2 keV. Mira A is toward the West (see Fig. 3); (b) contours of the ACIS-S
soft image (0.3-0.7keV) (toward the West) overlayed on the image of the hard
image (0.7-2 keV) (toward the East).}
\label{mira_xandhst}
\end{figure}
\clearpage

\newpage

\clearpage

\begin{figure}[t]
\centering
\scalebox{0.830}{\rotatebox{0}{\includegraphics{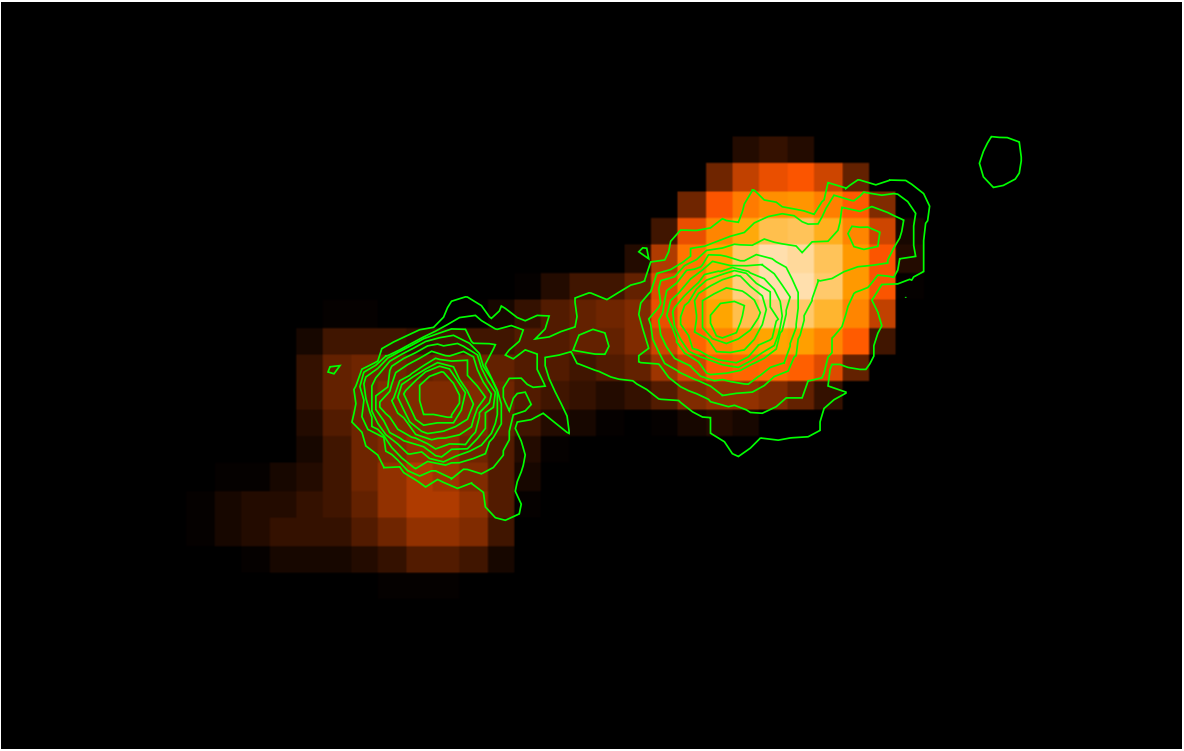}}}
\caption{{\it Chandra} image of Mira B (left) and Mira A (right),
separated by $\sim$ 0.6'', with overlayed
contours of the HST 3729 $\AA$ image of the system. North is up, East
is to the left. The apparent
extended point-like structures in the HST contours of Mira A in the NW
direction are PSF structures due to a red leak in the filter.}
\label{mira_xandhst}
\end{figure}

\clearpage

\end{document}